\documentclass[article]{revtex4}
\usepackage{graphicx}
\usepackage{amssymb}
\usepackage{bm}
\usepackage{hyperref}
\usepackage{epstopdf}
\usepackage{color}

\begin{document}

\title{Energy transition in vacuum matter coupling}

\author{Jiaxin Wang $^{1}$}
\author{Xinhe Meng$^{1, 2}$}

\affiliation{$^{1}$Department of Physics, Nankai University, Tianjin
300071, P.R.China}
\affiliation{$^{2}$ Kavli
Institute of Theoretical Physics China,\\CAS, Beijing 100190,
China.\vspace{1cm}}
\date{Nov. 30th 2013}
%
\begin{abstract}
The lately proposed vacuum matter coupling model (arXiv:1303.6568) seems to be able to provide solutions to a variety of problems in cosmology. This coupling expressed by a transition term between vacuum energy and matter, but we point out that an additional energy transition mechanism is necessary for further development of this model in order to explain observations.
\end{abstract}

\maketitle

We are interested in a recent work about proposing dark-energy and matter coupled cosmology~\cite{tpli}, which contains matter and vacuum energy as two main components at low redshift. The main idea was proposed as
\begin{equation}
\rho = \rho_0 a^3 = \rho_m + \rho_\lambda ,
\end{equation}
where the subscript $m$ represents energy density of matter, while the $\lambda$ stands for vacuum energy.

The Friedmann like equations of this model was expressed as
\begin{eqnarray}
\dot{a}^2 &=& \epsilon + \frac{\kappa^2}{3}\rho_d a^2, \label{f1}\\
\ddot{a} &=& -\frac{\kappa^2}{2R_0\rho_0}\rho_d a^3, \label{f2}
\end{eqnarray}
where $a$, the conventional scale factor; $\kappa^2 = 8\pi G$; $\epsilon$, a constant carrying the information of conservation of mechanical energy; $\rho_d = \rho_m-\rho_\lambda$, the differential energy density.

Notice that the above equations are not derived from modified Einstein equation, but through a classic way in flat space with homogeneous and isotropic principle complied~\cite{tpli}.
These Friedmann like equations are incompatible with Einstein equation in the frame of general relativity, but alternatively, 
we would like to test whether it can well explain the basic features of cosmological expansion.

The background evolution can be directly derived from the Friedmann like equations(~\ref{f1}-\ref{f2}), which is critical for model constraining and for revealing the energy transition between vacuum energy and matter inferred by the model itself. The first step
is re-express the Friedmann like equations in a dimensionless way
\begin{eqnarray}
{\rm H}^2 &=& {\rm H}_0^2 [\Omega_d + \Omega_{\epsilon 0}(1+z)^2], \label{mf1}\\
\ddot{a} &=& -\frac{\mu}{2}\Omega_d a^3, \label{mf2}
\end{eqnarray}
with the definitions read:
\begin{eqnarray}
{\rm H}_0^2 &=& \frac{\kappa^2}{3}\rho_c, \\
\Omega_d &=& \frac{\rho_d}{\rho_c},\\
\Omega_\epsilon &=& \frac{3\epsilon}{\kappa^2\rho_c a^2},\\
\Omega_{\epsilon 0} &=& \frac{3\epsilon}{\kappa^2 \rho_c},\\
\mu &=& \frac{\kappa^2 \rho_c}{R_0 \rho_0}.
\end{eqnarray}

Then we combine Eq.~\ref{mf1} and Eq.~\ref{mf2} in order to eliminating $\ddot{a}$.
The the redshift dependency of the differential energy density reads
\begin{equation}
\frac{\partial \Omega_d}{\partial z} = \Omega_d [\frac{2}{(1+z)} + \frac{\mu}{{\rm H}_0^2 (1+z)^3}], \label{d}
\end{equation}
whose general analytical solution can be written as
\begin{equation}
\Omega_d(z) = C (1+z)^2\exp{\{-\frac{\mu}{2{\rm H}_0^2}(1+z)^{-2}\}}, \label{gsol}
\end{equation}
where $C$ as a constant of integration, which should be defined by initial condition. This equation provides the explicit transition form between vacuum energy and matter, which is naturally generated from the coupling model itself. In the next section we will mention another possible form of energy transition is required in order to calibrate the model with observations.

With the redshift evolution function of differential energy density, the redshift dependency of the Hubble parameter reads
\begin{equation}
{\rm H}^2(z) = {\rm H}_0^2 [\Omega_d(z) + \Omega_{\epsilon 0} (1+z)^2].
\end{equation}

Notice that in Eq.~(\ref{d}), when $\Omega_d = 0$, the system reaches a steady state and the cosmological expansion rate is constant. When the constant of integration is taken positive, the cosmic expansion is decelerated; while the constant $C$ is negative, the cosmic expansion will be accelerated. According to the mathematical feature of Eq.~(\ref{gsol}), the three types expansion are independent of each other; which means without some kind of mechanism, one expansion type is locked for the whole universe history once it is admitted by observations. Only by allowing extra energy transition between the matter and vacuum energy powered by some kind of mechanism, the late-stage cosmological expansion could be accelerated or decelerated. We emphasize that there are two types of energy transition between matter and vacuum energy; one is the transition mechanism which is responsible for Eq.~(\ref{d}), the other mechanism is responsible for carrying $\Omega_d$ to cross above or below zero. In our present work, we mainly focus on discussing the second type of transition mechanism, so the energy transition mentioned below only represents the latter one.

It is obvious that Eq.~(\ref{gsol}), whose role is to reduce the difference between matter and vacuum density, does not have the ability to realize the transition phenomenon, thus additional transition mechanism is necessary for the phenomenological completeness of $\Lambda$DMC cosmology. To illustrate this, we show the evolution of $\Omega_d$ of $\Lambda$DMC with best-fit parameters given by OHD and $\Lambda$CDM in Fig~\ref{fig1}.

Restricted by Eq.~(\ref{d}), $\Omega_d$ should not rely only on redshift, but at least another redshift independent factor which can explain the change in the two coupled energy densities.
Let us assume that there exists one factor $\theta$ which is responsible for energy transition, then the differential energy density should reads $\Omega_d(z,\theta)$. Under this circumstance, the previous proposed interaction forms between dark matter and dark energy, which are mostly of the first type of transition mentioned above, are not adoptable, ie., general coupling between dark sectors~\cite{couple} or the vacuum decay~\cite{vacuumd}, where the energy transition strongly relies on energy density evolution. But they are available for explaining the mechanism which is responsible for Eq.~(\ref{d}). It may be possible that some particular forms of first type of transition can also provide explanation for the mechanism of the second one.

\begin{figure}[ht]
\begin{center}
\includegraphics[width=0.3\textwidth]{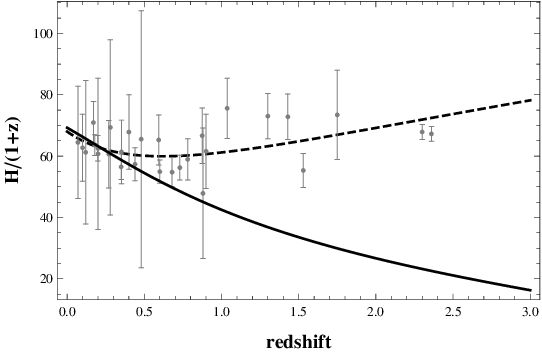}\\
\includegraphics[width=0.3\textwidth]{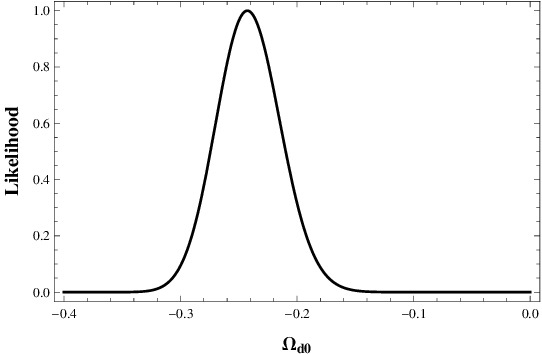}
\includegraphics[width=0.3\textwidth]{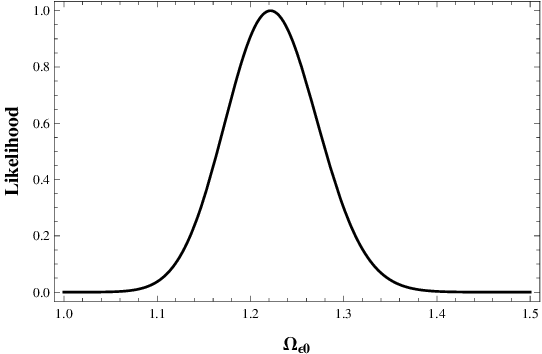}
\includegraphics[width=0.3\textwidth]{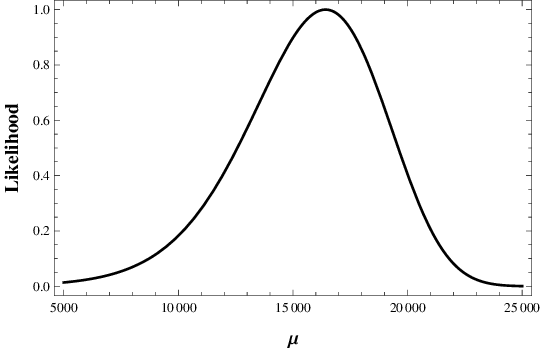}

\end{center}
\caption{{\it Upper}: The evolution of expansion rate with respect to redshift of $\Lambda$DMC model (solid line, best-fit parameters given by OHD, without energy transition mechanism) and $\Lambda$CDM model (dashed line, with $\Omega_{m0}=0.32$ and $\Omega_\Lambda=0.68$). {\it Lower}: The likelihood of parameters of $\Lambda$DMC model constrained by OHD. We test the distribution of parameters in redshift bin $[0,0.4]$, $[0,0.5]$, $[0,0.7]$ and even $[0,1.3]$ which offers similar results. }
\label{fig2}
\end{figure}

\begin{figure}[ht]
\begin{center}
\includegraphics[width=0.3\textwidth]{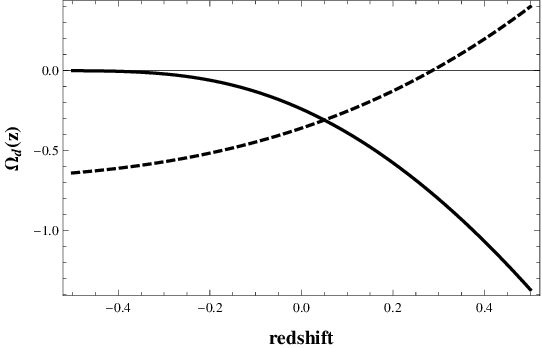}
\end{center}
\caption{The differential energy density of $\Lambda$DMC mdoel (solid line) and $\Lambda$CDM model (dashed line, with $\Omega_{m0}=0.32$ and $\Omega_\Lambda=0.68$). The value of parameters in $\Lambda$DMC is given by model fitting with OHD (see Fig.~\ref{fig2}). It is clear that if without energy transition mechanism, the universe is dominated by vacuum energy in the past thus large-scale structures cannot be formed.}
\label{fig1}
\end{figure}

Some low redsfhit astrophysical observations, ie., SNe, BAO and Observational Hubble parameter Data (OHD), are available for model test. The coupling model is constrained by OHD at low redshift ($z<1$) without assuming any kind of energy transition, the likelihood of distribution of parameters are shown in Fig.~\ref{fig2}.
For constraining model without energy transition by OHD, we use the expression of the differential energy density which reads
\begin{equation}
\Omega_d(z) = \Omega_{d0}(1+z)^2\exp{\{\frac{\mu}{2{\rm H}_0^2}[1-\frac{1}{(1+z)^2}]\}}.
\end{equation}

We find that at $2\sigma$ confidence level the acceleration is confirmed in the coupling model at present (since $\Omega_{d0}$ is negative at $2\sigma$ level), which is in consistent with the standard understanding of cosmology. Different redshift bins of OHD is adopted with ${\rm H}_0 = 70 km\cdot s^{-1}\cdot Mpc^{-1}$, the results about the parameters are robust with similar shapes of likelihoods and the best-fit values.

From Fig.~(\ref{fig1}) and the upper left panel in Fig.~(\ref{fig2}), we can conclude that without energy transition, the coupling model is in conflict against observation, besides the diversion with OHD at higher redshift, large-scale structure in our universe today can not be observed with vacuum energy dominated epoch in the past. $\Omega_d$ must be positive at high redshfit in order to allow the structure formation, thus the energy transition mechanism is necessary for the validity of the coupling model.

For BAO observation we have an additional free parameter $\Omega_0$, which comes from the following derivation:

Since we have
\begin{eqnarray}
\frac{\rho_d}{\rho_c} &=& \frac{\rho_m}{\rho_c} - \frac{\rho_\lambda}{\rho_c} = \Omega_d,\\
\frac{\rho}{\rho_c} &=& \frac{\rho_m}{\rho_c} + \frac{\rho_\lambda}{\rho_c} = \Omega_0(1+z)^3,
\end{eqnarray}
with the definitions
\begin{eqnarray}
\Omega_m &=& \frac{\rho_m}{\rho_c},\\
\Omega_\lambda &=& \frac{\rho_\lambda}{\rho_c},\\
\Omega_0 &=& \frac{\rho_0}{\rho_c}.
\end{eqnarray}
So the redshift dependency of matter and vacuum energy can be explicitly depicted as
\begin{eqnarray}
\Omega_m(z) &=& \frac{\Omega_d(z) + \Omega_0(1+z)^3}{2},\\
\Omega_\lambda(z) &=& \frac{\Omega_0(1+z)^3-\Omega_d(z)}{2}.
\end{eqnarray}
The dimensionless matter density $\Omega_{m0}$ appeared in the BAO parameter $\mathcal{A}$ can be expressed as
\begin{equation}
\Omega_{m0} = \frac{\Omega_{d0}+\Omega_0}{2\Omega_0},
\end{equation}
since the definition of dimensionless energy densities in this article is different from the conventional way, we should be careful when applying them in model constraining.

We combined the data list given in~\cite{h0} with the latest data obtained from BOSS DR11~\cite{h8}) as the OHD data-set; in terms of BAO, we combined the released data from SDSS, WiggleZ and 6dFGRS (\cite{bao1,bao2}); Union2.1 (\cite{sn}) set is adopted as SNe data-set. CMB priors or CMB anisotropic power spectrum are usable here after radiation energy is included in this model.

\end{document}